\documentclass[10pt,conference,compsocconf,letterpaper]{IEEEtran}
\newtheorem{theorem}{Theorem}

\newtheorem{lemma}[theorem]{Lemma}

\newtheorem{definition}{Definition}

\usepackage{cite}

\ifCLASSINFOpdf
\usepackage[pdftex]{graphicx}
\usepackage{subfigure}
\graphicspath{{../pdf/}{../jpeg/}}
   \DeclareGraphicsExtensions{.pdf,.jpeg,.png}
\else
\fi
\usepackage{amsmath}

\usepackage[ruled,vlined]{algorithm2e}
\usepackage{url}

\begin{document}

\title{Parallel Algorithms for Core Maintenance in Dynamic Graphs}

\author{\IEEEauthorblockN{
Na Wang, Dongxiao Yu, Hai Jin, Chen Qian, Xia Xie, Qiang-Sheng Hua\\
\IEEEauthorblockA{Services Computing Technology and System Lab\\Big Data Tecgnology and System Lab\\Clusters and Grid Computing Lab\\School of Computer\ Science and Technology.\\
Huazhong University of Science and Technology, Wuhan, 430074,\ China\\
Email: \{Ice$\_$lemon,dxyu,hjin,M201572720,shelicy,qshua\}@hust.edu.cn}
}}

\maketitle

\begin{abstract}
This paper initiates the studies of parallel algorithms for core maintenance in dynamic graphs. The core number is a fundamental index reflecting the cohesiveness of a graph, which are widely used in large-scale graph analytics. The core maintenance problem requires to update the core numbers of vertices after a set of edges and vertices are inserted into or deleted from the graph.  
We investigate the parallelism in the core update process when multiple edges and vertices are inserted or deleted. Specifically, we discover a structure called \emph{superior edge set}, the insertion or deletion of edges in which can be processed in parallel. 
Based on the structure of superior edge set, efficient parallel algorithms are then devised for incremental and decremental core maintenance respectively. To the best of our knowledge, the proposed algorithms are the first parallel ones for the fundamental core maintenance problem. The algorithms show a significant speedup in the processing time compared with previous results that sequentially handle edge and vertex insertions/deletions. 
Finally, extensive experiments are conducted on different types of real-world and synthetic datasets, and the results illustrate the efficiency, stability and scalability of the proposed algorithms.
\end{abstract}


%
\IEEEpeerreviewmaketitle

\section{Introduction}
Graph analytics has drawn much attention from research and industry communities, due to the wide applications of graph data in different domains. One of the major issues in graph analytics is identifying cohesive subgraphs. There are lots of indexes to depict the cohesiveness of a graph, such as cliques, k-truss, k-core, F-groups, n-clans and so on \cite{IntroToSN}, among which $k$-core is recognized as one of the most efficient and helpful one. Given a graph $G$, the $k$-core is the largest subgraph in $G$, such that the minimum degree of the subgraph is at least $k$. The core number of a vertex $v$ is defined as the largest $k$ such that there exists a $k$-core containing $v$. In static graphs, the computation of the core number of each vertex is known as the $k$-core decomposition problem. Besides the analysis of cohesive subgroup, $k$-core decomposition are widely used in a large number of applications to analyze the structure and function of a network. For example, the k-core decomposition can be used to analyze the topological structure of Internet \cite{app1}, \cite{app2}, to identify influential spreader in complex networks \cite{app3} \cite{app4}, to analyze the structure of large-scale software systems \cite{app5}\cite{app6}\cite{app7}\cite{app8}, to predict the function of biology network \cite{app9}, and to visualize large networks \cite{app10}\cite{app11} and so on.

In static graphs, the $k$-core decomposition problem has been well studied. The algorithm presented in \cite{O(m)} can compute the core number of each vertex in $O(m)$ time, where $m$ is the number of edges in the graph. However, in many real-world applications, graphs are subject to continuous changes like insertion or deletion of vertices and edges. In such dynamic graphs, many applications require to maintain the core number for
every vertex online, given the network changes over time. But it would be expensive to recompute the core numbers of vertices after every change of the graph, though the computation time is linear, as the size of the graph can be very large. Furthermore, the graph change may only affect the core numbers of a small part of vertices. Hence, the \emph{core maintenance} problem~\cite{Li2014TKDE} is recommended, which is to identify the vertices whose core numbers will be definitely changed and then update the core numbers of these vertices. There are two categories of core maintenance, \emph{incremental} and \emph{decremental}, which handle edge/vertex insertion and deletion respectively.

\begin{figure}[!t]\label{fig1}
\centering
\includegraphics[width=2.0in]{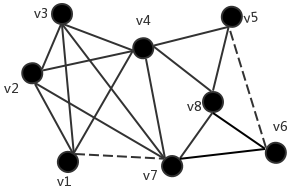}
\caption{Assume edge $<v5,v6>$ and $<v1,v7>$ are inserted. \textbf{TRAVERSAL} algorithm in \cite{Sar2016Incremental} processes them one by one. First for edge $<v5,v6>$, it will visit vertices $v5,v6,v8$ and update their core numbers. And then when inserting $<v1,v7>$, it will visit $v1,v2,v3,v4,v5,v6,v7,v8$, and update core numbers of $v1,v2,v3,v4,v7$. However in our parallel algorithm, edges $<v5,v6>$ and $<v1,v7>$ are handled in parallel using two processes. In the process handling $<v5,v6>$, the algorithm execution will visit and update $v5,v6,v8$, and in another process for $<v1,v7>$, the algorithm will visit and update $v1,v2,v3,v4,v7$. Hence, the parallel algorithm avoids duplicate visiting of $v_5,v_6,v_8$.}
\end{figure}

Previous works focus on maintaining the core numbers of vertices in the scenario that a single edge is inserted or deleted from the graph. For multiple edge/vertex insertions/deletions, the inserted/deleted edges are processed sequentially. The sequential processing approach, on the one hand, incurs extra overheads when multiple edges/vertices are inserted/deleted, as shown in Fig. \ref{fig1}, and on the hand, it does not fully make use of the computation power provided by multicore and distributed systems. Therefore, it is necessary to investigate the parallelism in the edge/vertex processing procedure and devise parallel algorithm that suits to implement in multicore and distributed systems. But to the best of our knowledge, there are no known parallel algorithms proposed for the core maintenance problem.

In the core maintenance problem, the insertions/deletions of vertices can be handled by implementing an edge insertion/deletion algorithm. Specifically, inserting a vertex is equivalent to the following process: first inserting the vertex into the graph by setting its core number as 0, and then inserting the edges connected to the new vertex. Similarly, the deletion of a vertex is equivalent to the process that deleting the edges connected to the vertex and finally deleting the vertex. Hence, in this paper, we only consider the edge insertions and deletions.  

It is a very difficult task to design parallel algorithms for core maintenance in dynamic graphs. Different from the single edge insertion/deletion case, where the core number of each vertex changes by at most 1, it is hard to identify the change of a vertex's core number in the multiple edge insertion/deletion scenario, as the change of a vertex' core number may be affected by several inserted edges. An intuitive manner is to split the inserted/deleted edges into sets that affect disjoint sets of vertices in the original graph. However, the parallelism of this manner is poor. In this work, we take a more efficient approach that exhibits better parallelism. Specifically, we propose a structure called \emph{superior edge set}. The inserted/deleted edges can be split into multiple superior edges sets, and for each vertex connected to inserted/deleted edges, a superior edge set contains at least one inserted edge connected to it. It is shown that the insertion or deletion of edges in a superior edge set can change the core number of every vertex by at most 1. Hence, the core numbers of vertices when inserting or deleting a superior edge set can be maintained using a parallel procedure: first identifying the vertices whose core numbers will change due to the insertion or deletion of every edge in parallel, and then updating the core number of these vertices by 1. A parallel algorithm can then be obtained by iteratively handling the insertions/deletions of split superior edge sets using the above parallel procedure.

In summary, our contributions are summarized as follows.
\begin{itemize}
\item We propose a structure called \emph{superior edge set}, and show that if the edges of a superior edge set is inserted into/deleted from a graph, the core number of each vertex can change by at most 1. It implies that the insertion/deletion of these edges can be processed in parallel. We also give sufficient conditions for identifying the vertices whose core numbers will change, when inserting/deleting a superior edge set.
\item We then present parallel algorithms for incremental and decremental core maintenance respectively. Comparing with sequential algorithms, our algorithms reduce the number of iterations for processing $s$ inserted/deleted edges from $s$ to the maximum number of edges inserted to each vertex. In large-scale graphs, the acceleration is significant, since each vertex can connect to only a few inserted or deleted edges. For example, as shown in the experiments, even if inserting $2\times 10^4$ edges to the LiveJournal graph (refer to Table \ref{table_graph} in Section~\ref{sec:experiment}), the number of iterations is just 3 in our parallel algorithms, in contrast with $2\times 10^4$ ones in sequential processing algorithms. 
\end{itemize}
We also conduct extensive experiments over both real-world and synthetic graphs, to evaluate the efficiency, stability and scalability of our algorithms. The results show that comparing with sequential processing algorithms, our algorithms significantly speed up core maintenance, especially in cases of large-scale graphs and large amounts of edge insertions/deletions.


The rest of this paper is organized as follows. In Section~\ref{sec:relate}, we briefly review closely related works. In Section~\ref{sec:problem}, the problem definitions are given. Theoretical results supporting the algorithm design are presented in Section~\ref{sec:basis}. The incremental and decremental parallel algorithms are proposed in Section~\ref{sec:in} and Section~\ref{sec:de} respectively. In Section~\ref{sec:experiment}, the experiment results are illustrated and analyzed. The whole paper is concluded in Section~\ref{sec:conclusion}.


\section{Related Work}\label{sec:relate}

In static graphs, the core decomposition problem has been extensively studied. The state-of-the art algorithm was given in \cite{O(m)}, the runtime of which is linear in the number of edges. In \cite{massivenetwork}, an external-memory algorithm was proposed when the graph is too large to hold in memory. Core decomposition in the distributed setting was studied in \cite{TPDS13}. The above three algorithms were compared in \cite{SinglePC} under the GraphChi and WebGraph models. Parallel core decomposition was studied in \cite{ParK}. 


Core maintenance in dynamic graphs has also been widely studied. However, all previous works focus on the case of single edge insertion/deletion, and sequentially handle multiple edge insertions/deletions. Efficient algorithms were proposed in \cite{Sar2016Incremental,Li2014TKDE}. In \cite{I/OICDE16}, an algorithm was proposed to improve the I/O efficiency. Furthermore, \cite{AksuTKDE14} and \cite{DEBS16} solved the core maintenance problem in the distributed environment.


\section{Problem Definitions}\label{sec:problem}
We consider an undirected, unweighted simple graph $G = (V,E)$, where $V$ is the set of vertices and $E$ is the set of edges. Let $n = |V|$ and $m = |E|$. For a node $u\in V$, the set of its neighbors in $G$ is denoted as $N(u)$, i.e., $N(u) = \{v\in V | (v,u) \in E\}$. The number of $u$'s neighbors in $G$ is called the degree of $u$, denoted as $d_G(u)$. So $d_G(u) =|N(u)|$. The maximum and minimum degree of nodes in $G$ is denoted as $\Delta(G)$ and $\delta(G)$ respectively. We next give formal definitions for the \emph{core number} of a vertex and other related concepts.


\begin{definition}[\textbf{$k$-Core}]
Given a graph $G=(V,E)$ and an integer $k$, the $k$-core is a maximal connected subgraph $H$ of $G$, where each vertex has at least $k$ neighbors in $H$, i.e., $\delta(H)\ge k$.
\end{definition}

\begin{definition}[\textbf{Core Number}]
Given a graph $G=(V,E)$, the core number of a vertex $u\in G$, denoted by $core_G(u)$, is the the largest $k$, such that there exists a $k$-core containing $u$. For simplicity, we use $core(u)$ to denote $core_G(u)$ when the context is clear.
\end{definition}

\begin{definition}[\textbf{Max-k-Core}]
The max-k-core associated with a vertex $u$, denoted by $H_u$, is the $k$-core with $k=core(u)$.
\end{definition}


In this work, we aim at maintaining the core numbers of vertices in dynamic graphs. 
Specifically, we define two categories of graph changes: \emph{incremental}, where a set of edges $E'$ are inserted to the original graph, and \emph{decremental}, where a set of edges are deleted. Based on the above classification, we distinguish the core maintenance problem into two scenarios, as defined below.


%


\begin{definition}[\textbf{Incremental Core Maintenance}]
Given a graph $G=(V,E)$, the incremental core maintenance problem is to update the core numbers of vertices after an incremental change to $G$.
\end{definition}
\begin{definition}[\textbf{Decremental Core Maintenance}]
Given a graph $G=(V,E)$, the decremental core maintenance problem is to update the core numbers of vertices after a decremental change to $G$.
\end{definition}
\section{Thoeretical Basis}\label{sec:basis}

In this section, we give some theoretical Lemmas that constitute the theoretical basis of our algorithms.

At first, we introduce some definitions. Given a graph $G=(V,E)$, an edge $e=<u,v>$ is called a superior edge for $u$ if $core_G(v)\geq core_G(u)$. Notice that in the definition, we do not require $e\in E$, i.e., $e$ may be an edge that is about to insert to graph $G$. Furthermore, we define the core number of an edge as the smaller one of its endpoints, i.e., $core_G(e)=\min\{core_G(u),core_G(v)\}$.
 
\begin{definition}[\textbf{k-Superior Edge Set}]
An edge set $E_k=\{e_1,e_2,...,e_p\}$ is called an $k$-superior edge set, if for each edge $e_i=<u_i,v_i>,1\le i\le p$, it satisfies:

(i) $e_i$ is a superior edge with core number $k$.

(ii) if $e_i$ and $e_j (1\le i,j\le p, i\neq j)$ have an common endpoint $u'$, $core_G(u')>k$.
\end{definition}

In other words, in a $k$-superior edge set $E_k$, each edge is a superior edge for a vertex with core number $k$, and in $E_k$, each vertex connects to at most one superior edge for it. 

The union of several $k$-superior edge sets with distinct $k$ values is called a \emph{superior edge set}. It can be known that in a superior edge set, each vertex can still connect to at most one superior edge for it. 


In the following, we will first show that when inserting/deleting a superior edge set (Lemma \ref{Them:superioredgesetinsert} and Lemma \ref{Them:superioredgesetdelete}), the core number of every vertex can change by at most 1, and then give a sufficient condition for identifying vertices whose core numbers change (Lemma \ref{le:exkpaths}, Lemma \ref{corollary:csd} and Lemma \ref{corollary:sd}).

\subsection{Superior Edge Set Insertion/Deletion}


%
We first prove a result on the core number increase of every vertex when inserting a $k$-supeior edge set. For simplicity, we use $core(u)$ to denote $core_G(u)$ when the context is clear.

\begin{lemma}\label{Lem:k-superioredgesetinsert}
Given a graph $G=(V,E)$, if a $k$-superior edge set $E_k=\{e_1,e_2,e_3,...,e_p\}$ is inserted to $G$, where $k\geq 0$, for each node $v$, it holds that:\\
$(i)$ if $core(v)=k$, $core(v)$ can increase by at most 1;\\
$(ii)$ if $core(v)\neq k$, $core(v)$ will not change.
\end{lemma}

\begin{IEEEproof}
For $(i)$, we need to show that for a vetex $v$ with $core(v)= k$, $core(v)$ can increase by at most 1. Otherwise, assume $core(v)$ increases by $x$ to $k+x$, where $x>1$. Let $H_v$ and $H^+_v$ be the max-$k$-core of $v$ before edge insertion and the max-$(k+x)$-core of $v$ after edge insertion respectively. Then, $\delta(H_v)=k, \delta(H^+_v)=k+x$. It can be concluded that one of inserted edges must belong to $H^+_v$, as otherwise $core(v)=k+x$ before insertion as well. Let $Z=H^+_v\setminus E_k$. For a vertex $u\in Z$, if $core_G(u)<k$, the degree of $u$ does not change when deleting the edges in $E_k$, so $d_Z(u)\ge k+x$. If $core_G(u)=k$, $u$ can lose at most one neighbor that is connected by a superior edge for it in $E_k$, so $d_Z(u)\ge k+x-1$. If $core_G(u)\ge k+1$, $u$ must have at least $k$+1 neighbors whose core numbers are not smaller than $k$+1 in $G$. We add the vertices whose $core_G$ is larger than $k$ back to $Z$, and denote the induced graph as $Z'$. It can be obtained that $Z\subseteq Z'\subseteq G$. And from $G$ to $Z'$, $u$ does not lose any neighbor whose $core_G$ is not smaller than $k$+1. Hence, in $Z'$, $d_{Z'}(u)\geq k+1$. Then it can be seen that each vertex in $Z'$ has a degree at least $k$+1, i.e., $\delta(Z')\ge k+1$. This means that $core_{Z'}(v)\ge \delta(Z')>k$, which contradicts with $core_{Z'}(v)\leq core_G(v)=k$. Hence, $core(v)$ can increase by at most 1.

For $(ii)$, we need to show that for a vertex $v$ if $core(v)\neq k$, $core(v)$ cannot change. We consider two cases: $core(v) >k$ and $core(v) < k$. Assume $core(v)$ = $y$ increases by $x$ to $y+x$, where $x\ge 1$. Let $H_v$ and $H^+_v$ be the max-$y$-core of $v$ before edge insertion and the max-$(y+x)$-core after edge insertion respectively. Then we have $\delta(H_v)=y$, $\delta(H^+_v)=y+x$.

We first consider the $core(v) > $k$ $ case. There must be at least one of the edges $e_i$ in $E_k$ belonging to $H^+_v$, as otherwise $y=core(v)\geq\delta(H^+_v)=y+x$ before edge insertion.  Consider the edge $e_i$. At least one of its endpoints has a core number $k$, since $E_k$ is a $k$-superior edge set. Denote by $u'$ the endpoint of $e_i$ with core number $k$. As shown before, $core(u')$ can increase by at most 1. Hence, after the edge insertion, $core(u')\leq k+1<y+x$. This means that $u'$ is not in $H^+_v$, which is a contradiction. Therefore, if $core(v) > k$, $core(v)$ will not change after the edge insertion.

We next consider the $core(v) < k $ case. Similar as beofore, it can be shown that at least one of the edges $e_i=<u_i,v_i>$ in $E_k$ that is contained in $H^+_v$. Let $Z=H^+_v\setminus E_k$. Let $u$ be a vertex in $Z$, we consider three cases. If $core_G(u)=k$ and as proved before, it can be obtained that $core_Z(u)=k$. If $core_G(u)>k$, as shown before, the core number of $u$ will not be affected by the edge insertions. If $core_G(u)<k$, because $\delta(H^+_v)=y+x$, and $u$ does not connect to edges in $E_k$, we can get that $d_G(u) = d_Z(u)\geq y+x$. Let $s=min(k,y+x)$. Based on above, it can be obtained that $Z$ is a $s$-core and $core_Z(v)\ge s$. But this contradicts with the fact that $core_Z(v)\leq core_G(v)=y$. Then we can get that the core number of $v$ does not change after inserting $E_k$. 

Combining all above together, the Lemma is proved.
\end{IEEEproof}

Using a similar argument as that for proving Lemma \ref{Lem:k-superioredgesetinsert}, we can get that the core number changes of vertices after deleting a $k$-superior edge set from graph $G$, as given in the following Lemma.

\begin{lemma}\label{Lem:k-superioredgesetdelete}
Given a graph $G=(V,E)$, if a $k$-superior edge set $E_k=\{e_1,e_2,e_3,...,e_p\}$ is deleted from $G$, where $k\geq 0$, for each vertex $v$, it holds that:\\
$(i)$ if $core(v)=k$, $core(v)$ can decrease by at most 1;\\
$(ii)$ if $core(v)\neq k$, $core(v)$ will not change.
\end{lemma}

From the above Lemma \ref{Lem:k-superioredgesetinsert} and Lemma \ref{Lem:k-superioredgesetdelete}, we have known that for a graph $G=(V,E)$, after a $k$-superior edge set $E_k=\{e_1,e_2,e_3,...,e_p\}$ is inserted into or deleted from $G$, only vertices with core numbers $k$ may increase/decrease, and the change is at most 1. This implies that if a $k$-superior edge set is inserted/deleted, it will be enough to only visit vertices whose core numbers are $k$ and check if their core numbers will be updated. And because the core numbers of these vertices can change by at most 1, we can handle these edge insertions in parallel: first we find the update set of vertices that will change core numbers because of the insertion of each particular edge in parallel, and the union of these update sets is just the set of vertices whose core numbers will change by one. 



In fact, we can get even better results, which are given in the following Lemma~\ref{Them:superioredgesetinsert} and Lemma~\ref{Them:superioredgesetdelete}. 

\begin{lemma}\label{Them:superioredgesetinsert}
Given a graph $G=(V,E)$ and a superior edge set $\mathcal{E}_q = E_{k_1}\cup E_{k_2}\cup,...,\cup E_{k_q} $, where $E_{k_i}$ for $1\leq i\leq q$ is a $k_i$-superior edge set and $k_i<k_j$ if $i<j$, it holds that after inserting $\mathcal{E}_q$ into $G$, the core number of each vertex $u$ can increase by at most 1.
\end{lemma}

\begin{IEEEproof}
It can be seen that inserting edges in $\mathcal{E}_q$ into $G$ all together has the same result with inserting $E_{k_i}$ one by one. We next assume $E_{k_i}$ are inserted one by one. To prove the Lemma, we need to prove that if inserting $E_{k_i}$ makes a vertex increase its core number from $k_i$ to $k_i$+1, its core number cannot change any more when inserting $E_{k_{j}}$ for $j>i$. Clearly, we only need to prove the above result for $E_{k_{i+1}}$. There are two cases we need to consider.

If $k_{i+1} > k_i+1$, by Lemma~\ref{Lem:k-superioredgesetinsert}, the core number of $u$ will not increase any more when inserting $E_{k_{i+1}}$, since only vertices with core numbers of $k_{i+1}$ may increase their core numbers. 

We next consider the case of $k_{i+1} = k_{i} + 1$. We claim that if there is a vertex increasing its core number from $k_i$ to $k_i+2$ after the insertions of $E_{k_i}$ and $E_{k_{i+1}}$, the vertex must have a neighbor which increases the core number from $k_i$ to $k_i+2$ as well during the insertions. 
Let $u$ be a vertex whose core number is increased from $k_i$ to $k_i+2$ after inserting $E_{k_i}$ and $E_{k_{i+1}}$. Notice that $u$ does not connect to edges in $E_{k_{i+1}}$. Hence, the degree of $u$ does not change when inserting $E_{k_{i+1}}$. Furthermore, by Lemma~\ref{Lem:k-superioredgesetinsert}, the core number of each neighbor of $u$ can be increased by at most 1. So $u$ has at least $k_i+2$ neighbors whose core numbers are not smaller than $k_i+1$ and some of these neighbors have a core number of $k_i+1$. Denote by $P_{k_i+1}(u)$ the vertices in $N(u)$ whose core numbers are $k_i+1$ before inserting $E_{k_{i+1}}$. It can be obtained that there must be a vertex $w\in P_{k_i+1}(u)$ whose core number is $k_i$ before inserting $E_{k_i}$, as otherwise, the core number of $u$ is $k_i+1$ before inserting $E_{k_i}$, which contradicts with our assumption.

Let $V_2$ denote the set of vertices whose core numbers change from $k_i$ to $k_i+2$ after the insertions of $E_{k_i}$ and $E_{k_{i+1}}$. Because inserting $E_{k_{i+1}}$ does not change the degrees of vertices in $V_2$, there must be a vertex $w\in V_2$ whose core number change is caused because of the core number change of vertices in $N(w)\setminus V_2$, as otherwise no vertex in $V_2$ can change the core number. Let $w'$ be a vertex in $N(w)\setminus V_2$ whose core number change causes the core number change of $w$. Then $core(w')$ is $k_i+1$ before inserting $E_{k_i}$ and is increased to $k_i+2$ after inserting $E_{k_{i+1}}$. To make $w$ increase its core number to $k_i+2$ after inserting $E_{k_{i+1}}$, there must be at least $k_i+2$ neighbors in $N(w)\setminus V_2$ whose core numbers are initially not smaller than $k_i+1$ before inserting $E_{k_i}$ and $E_{k_{i+1}}$. It concludes that $core(w) = k_i+1$ before inserting $E_{k_i}$ and $E_{k_{i+1}}$. However, this contradicts with the fact that $core(w)$ is $k_i$ before insertions. The contradiction shows that if the core number of a vertex is changed when inserting $E_{k_i}$, its core number will not change any more when inserting $E_{k_{i+1}}$. 

Combining all above together, the Lemma is prove.
\end{IEEEproof}

Similarly, for the case of a superior edge set deletion, we have the following result.

\begin{lemma}\label{Them:superioredgesetdelete}
Given a graph $G=(V,E)$ and a superior edge set $\mathcal{E}_q = E_{k_1}\cup E_{k_2}\cup,...,\cup E_{k_q}$, where $E_{k_i}$ for $1\leq i\leq q$ is a $k_i$-superior edge set and $k_i<k_j$ if $i<j$, it holds that after deleting $\mathcal{E}_q$ from $G$, the core number of each vertex $u$ can decrease by at most 1.
\end{lemma}

In above, we have shown that when inserting or deleting a superior edge set from a graph, the core numbers of vertices can change by at most 1. This implies that the core updates of inserting/deleting edges in a superior edge set can be processed in parallel by distributing distinct $k$-superior edge sets to distinct processes. Furthermore, we have also shown which set of vertices may change due to the insertion or deletion of a $k$-superior edge set. In the subsequent section, we give more accurate conditions for a vertex to change its core number when inserting/deleting a superior edge set.

\subsection{Core Number Change}
We first introduce some notations.
\begin{definition}[\textbf{Superior Degree}]\label{de:sd}
For a vertex $u$ in a graph $G$, $v$ is a \emph{superior neighbor} of $u$ if the edge $<u,v>$ is a \emph{superior edge} of $u$. The number of $u$'s superior neighbors is called the \emph{superior degree} of $u$, denoted as $SD(u)$.
\end{definition}

It can be known that only superior neighbors of a vertex may affect the change of its core number. 


\begin{definition}[\textbf{Constraint Superior Degree}]
The \emph{constraint superior degree} $CSD(u)$ of a vertex $u$ is the number of $u$'s neighbors $w$ that satisfies $core(w) > core(u)$ or $core(w) = core(u) \land SD(w) > core(u) $. 
\end{definition}

For a vertex $u$, its constraint superior degree $CSD(u)$ is the number of $u$' neighbors $w$, that has a larger core number than $u$ or has the same core number but has enough neighbors to support itself to increase core number. 

\begin{definition}[\textbf{K-Path-Tree}]\label{Def:KPT_u}
For a vertex $u$ with a core number $core(u)$, the \emph{$K$-Path-Tree} of $u$ is a DFS tree rooted at $u$ and each vertex $w$ in the tree satisfies $core(w) = core(u)$. For simplicity we use $KPT_u$ to represent K-Path-Tree of $u$.
\end{definition}

The \textbf{$KPT_u$} includes all vertices $w$ with $core(w) = core(u)$ that are reachable from $u$ via paths that consists of vertices with core numbers equal to $core(u)$. When a superior edge of $u$ is inserted or deleted, as shown in Lemma \ref{Lem:k-superioredgesetinsert}, only vertices in $KPT_u$ may change their core numbers. And for the insertion case, a more accurate condition was given in \cite{Sar2016Incremental} for identifying the set of vertices that may change core numbers, as shown below.

\begin{lemma}\label{Them:kpath}
Given a graph $G=(V,E)$, if an edge $<u,v>$ is inserted and $core(u)\le core(v)$, then only vertices $w$ in the \textbf{$KPT_u$} of $u$ and $CSD(w) > core(u)$ may have their core numbers increased, and the increase is no more than 1.
\end{lemma}

However, the above Lemma~\ref{Them:kpath} is just suitable for the one edge insertion scenario. We next generalize the above result to the scenario of inserting a $k$-superior edge set, as shown in Lemma~\ref{Them:exK-Path} below, which will help find the set of vertices with core number changes when inserting multiple edges. Before giving the result, we need to generalize the concept of $K$-path-tree to $exK$-path-tree.

\begin{definition}[\textbf{$exK$-Path-tree}]\label{Def:exK-Path-Tree}
For a $k$-superior edge set $E_k$ = \{$e_1,e_2,...,e_p$\}, w.l.o.g., assume that for each $e_i$ = $<u_i,v_i>$, $core(v_i) \ge core(u_i) = k$. The union of $KPT_{u_i}$ for every $u_i$ is called the \emph{exK-path-tree} of $E_k$. For simplicity we use $exKPT$ to represent exK-path-tree of $E_k$. 
\end{definition}


By Lemma~\ref{Them:kpath}, we can get that when inserting a $k$-superior edge set $E_k$, only vertices $w$ in the $exKPT$ satisfying $CSD(w)> k$ may have their core numbers change, and Lemma \ref{Lem:k-superioredgesetinsert} ensures that these vertices can change their core numbers by at most 1. This result is summarized in the following Lemma.

\begin{lemma}\label{Them:exK-Path}
Given a graph $G=(V,E)$, if a $k$-superior edge set $E_k$ is inserted, then only vertices $w$ in the $exKPT$ satisfying $CSD(w)> k$ may have their core numbers increased, and the core change is at most 1.
\end{lemma}

The above Lemma \ref{Them:exK-Path} implies that after an edge in a $k$-superior edge set $E_k$ is inserted, the vertices whose core numbers change during the insertion will not change any more when inserting other edges in $E_k$. Based on the above result and Lemma \ref{Them:superioredgesetinsert} and Lemma \ref{Them:superioredgesetdelete}, we can get the set of vertices whose core number change when inserting a superior edge set.

\begin{lemma}\label{le:exkpaths}
Given a graph $G=(V,E)$, if a superior edge set $\mathcal{E}_q=E_{k_1}\cup\ldots\cup E_{k_q}$ is inserted, then only vertices $w$ in every \textbf{$exK_iPT$}s of every $E_{k_i}$ for $1\leq i\leq q$ satisfying $CSD(w)> k$ may have their core numbers increased, and the core number change can be at most 1.
\end{lemma}

By the definition of $CSD$, we have the following result.
\begin{lemma}\label{corollary:csd}
Given a graph $G=(V,E)$ and a vertex $u$ with core number $k$. After inserting a superior edge set into $G$, if $CSD(u) \le k$, then $u$ cannot be in a $k+1$-core. 
\end{lemma}

Lemma~\ref{le:exkpaths} and Lemma \ref{corollary:csd} give accurate conditions to determine the set of vertices that will change the core numbers, after inserting a superior edge set. 

For deletion case, we have the following result, which can be obtained directly from Definition \ref{de:sd}.

\begin{lemma}\label{corollary:sd}
After deleting an edge set from a graph $G=(V,E)$, for $u\in V$, if $core(u)$ = $k$ and $SD(u) \le k$, then $u$ will decrease its core number. 
\end{lemma}

In this section, we have given accurate conditions for the core number changes of vertices after inserting/deleting a superior edge set. In the subsequent Section~\ref{sec:in} and Section~\ref{sec:de}, we will show how to utilize these theoretical results to design parallel algorithms for incremental and decremental core maintenance respectively.

\section{Incremental Core Maintenance}\label{sec:in}

In this section, we present the algorithm for incremental core maintenance, whose pseudo-code is given in Algorithm~\ref{Alg:SuperiorEdgeInsert}. We consider the core number update of vertices after inserting a set of edges $E'$ to graph $G=(V,E)$. Let $V'$ denote the set of vertices connecting to edges in $E'$. The set of core numbers of vertices in $V'$ is denoted as $\mathcal{C}$.

\begin{algorithm} [htb]\label{Alg:SuperiorEdgeInsert}  
\caption{SuperiorEdgeInsert($G,E',V',core()$)}
\textbf{Input}\\
The graph, $G=(V,E)$;\\
The inserted edge set, $E'$;\\
The set of vertices $V'$ connected to edges in $E'$\;
The core number $core(v)$ of each vertex in $V$;\\
\While{$E'$ is not empty}{
\nl \For{each vertex $u$ in $V'$}{
\nl   \If{$u$ connects a superior edge in $E'$ and $core(u)\notin \mathcal{C}$}{
     add core($u$) to $\mathcal{C}$\;
        }
    }
\nl \For{each core number $k$ in $\mathcal{C}$ in parallel} {
    $E_k\gets${ComputeSuperiorEdgeSet($k$)}\;
}
\nl insert $\cup_{k\in\mathcal{C}}E_k$ into $G$\; 
\nl delete $\cup_{k\in\mathcal{C}}E_k$ from $E'$\;
\nl     \For{each core number $k$ in $\mathcal{C}$ in parallel}{
     $V_k\gets${$K$-SuperiorInsert($G$,$E_k$,$core()$)}\;
    }
\nl \For{each vertex $v$ in $\cup_{k\in\mathcal{C}}V_k$} {
        $core(v) \gets{core(v)+1}$\;       
    }
}
\end{algorithm} 

\begin{algorithm} [htb]\label{Alg:FindKSuperiorEdges}  
\caption{ComputeSuperiorEdgeSet($k$)}
\textbf{Input}\\
The graph, $G=(V,E)$;\\
The update edge set, $E'$;\\
The set of vertices $V'$ connected to edges in $E'$\;
A core number $k$;\\
\nl $E_k\gets{\emptyset}$\;
\nl \For{each vertex in $V'$ with core number $k$}{
\nl         find a superior edge $<u,v>$ of $u$ from $E'$\;
\nl         add $<u,v>$ to $E_k$\;
    }
\nl \textbf{return} $E_k$\;
\end{algorithm} 


\begin{algorithm}  [htb]\label{Alg:insertK}
\caption{$K$-SuperiorInsert($G, E_k,$ core())}     
\textbf{Input}\\
The graph, $G=(V,E)$;\\
The $k$-superior edge set, $E_k$;\\
The core number $core(v)$ of each vertex in $V$\;
\textbf{Initially}, {$S\gets$ empty stack}\;
for each vertex $v\in V$, $visited[v]\gets{false}, removed[v]\gets{false}, cd[v]\gets{0}$\;
\nl compute $SD(v)$ for each vertex $v$ in $exKPT$ of $E_k$\;

\nl \For{each $e_i=<u_i,v_i> \in E_k$}{  
\nl \lIf{$core(u_i)\geq core(v_i)$}{
$r\gets{v_i}$}
\lElse{$r\gets{u_i}$}
\nl \If{visited[$r$] = false and removed[$r$] = false}{
\nl     \lIf{$CSD[r] =$ 0} {compute $CSD[r]$}
        \lIf{$cd[r] >=$ 0} {$cd[r]\gets{CSD[r]}$}
    \lElse
        {$cd[r]\gets{cd[r]+CSD[r]}$}
        $S.push(r)$\;
        $visited[r]\gets{true}$\;            
\nl     \While{S is not empty} { 
            $v\gets{S.pop()}$\;
\nl         \If{$cd[v]> k$}{
\nl             \For{each $<v,w>\in E$}{
\nl                 \If{$core(w) = k$ and $SD(w) > k$ and $visited[w]$ = false}{
                       $S.push(w)$\;
                       $visited[w] \gets{ true}$\;
                       \If{$CSD[w] =$ 0} {compute $CSD[w]$}
                        $cd[w] \gets{ cd[w] + CSD[w]}$ 
                    }
                } 
             }
\nl         \Else {
                \If{removed[$v$]=false}{
							InsertRemove($G$,core(),$cd$[],removed[],$k,v$)
						}
            }  
        } 
    }
}
\nl \For{each vertex $v$ in $G$}{
    \If{removed[$v$]=false and visited[$v$] = true}{
         $V_k \gets{V_k\cup\{v\}}$
    }
}
\nl \textbf{return} $V_k$;
\end{algorithm}  

\begin{algorithm}[htb]\label{Alg:InsertRemove} 
\caption{InsertRemove($G$,core(),cd[],removed[],$k,r$)} 
\nl $S\gets{ empty\ stack}$\;
\nl $S.push(r)$\;
\nl $removed[r] \gets{true}$\;
\nl \While{S is not empty}{
        $v\gets{S.pop()}$\;
        \For{each $<v,w>\in E$}{
            \If{core($w$) = $k$}{
\nl             $cd[w]\gets{cd[w]-1}$\;
                \If{cd[$w$] = $k$ and removed[$w$] = false}{
\nl                 $S.push(w)$\;
\nl                 $removed[w] \gets{ true}$\;
                }        
            }
        }
    }
\end{algorithm} 
The algorithm is executed in iterations. Basically, the algorithm split the inserted edges into multiple superior edge sets, and process the insertion of one superior edge set in one iteration. In each iteration, it first uses a parallel algorithm to find a suporior edge set from the inserted edges that have not been processed so far (Line 3). Then a parallel algorithm is executed for each edge in parallel to identify the set of vertices whose core numbers change, and increase the core numbers of these vertices by 1 (Line 6-7). It deserves to point out that we do not use directly algorithms handling single edge insertion/deletion as subroutine. Instead, we make the edges inserted with the same core number processed together, as we find that this can efficiently avioding duplicate visiting of vertices, which further accelerates our parallel processing procedure. We next introduce the two parts in each iteration respectively. 

Because the superior edges of vertices with different core numbers are disjoint, the $k$-superior edge sets $\{E_k\}$ for different core numbers $k\in\mathcal{C}$ can be computed in parallel using Algorithm~\ref{Alg:FindKSuperiorEdges}. Then the computed superior edge set is inserted into the graph and deleted from $E'$. 

The set of vertices with core number changes is also computed in parallel. Specifically, for each $k\in \mathcal{C}$, a child process is assigned to find the vertices whose core number changes are caused by the insertion of the computed $k$-superior edge set, using Algorithm~\ref{Alg:insertK}. Algorithm \ref{Alg:insertK} first computes $SD$ values for each vertex in $exKPT$ of $E_k$, and then for each edge $e_i = <u_i,v_i>$ in a $k$-superior edge set, finds the set of vertices whose core numbers change due to the insertion of $e_i$. For $e_i$, a \emph{positive} Depth-First-Search (DFS) is conducted on vertices in $KPT_r$ from the root vertex $r$, which is one of $u_i$ or $v_i$ that has a core number $k$\footnote{If both $u_i$ and $v_i$ have a core number equal to $k$, then $r$ can be either $u_i$ or $v_i$.}, to explore the set of vertices whose core numbers potentially change. In the algorithm, the $cd$ value of each vertex $v$ is used to evaluate the potential of a vertex to increase its core number, which records the dynamic changes of $CSD$ value. The intial value of $cd(v)$ is set as $CSD(v)$. For a vertex $v$, if $cd[v]\le k$, its core number cannot increase. If a vertex $v$ with $cd[v]\le k$ is traversed in the positive DFS procedure, a \emph{negative} DFS procedure initiated from $v$ will be started, to remove $v$ and update the $cd$ values of other vertices with core number $k$. After all vertices in $KPT_r$ are traversed, the vertices that are visited but not removed will increase the core numbers by 1.


\textbf{Performance Analysis.}
We next analyze the correctness and efficiency of the proposed incremental algorithm. At first, some notations are defined, which will be used in measuring the time complexity of the algorithm.

For graph $G=(V,E)$, the inserted edge set $E'$ and a subset $S$ of $E'$, let $G_S=(V,E\cup S)$ and $K(G_S)$ be the set of core numbers of vertices in $G_S$.

For $G_S$, let $L_S = \max_{u\in V}\{CSD(u)-core_{G_S}(u),0\}$. As shown later, $L_S$ is the max times a vertex $u$ can be visited by negative DFS procedures in the algorithm execution.

For $k\in K(G_S)$, let $V_S(k)$ be the set of vertices with core number $k$, and $N(V_S(k))$ be the neighbors of vertices in $V_S(k)$. Let $n_S=\max\{|V_S(k)|: k\in K(G_S)\}$.

Denoted by $E[V_S(k)]$ the set of edges in $G_S$ that are connected to vertices in $V_S(k)\cup N(V_S(k))$. Then we define $m_S$ as follows, which represents the max number of edges travelled when computing $SD$ in the case of inserting edges to $G_S$. 
\begin{equation*}
m_S=\max_{k\in K(G_S)}\{|E[V_S(k)]|\}.
\end{equation*}

Furthermore, we define the \emph{maximum inserted degree} as the maximum number of edges inserted to each vertex in $V$, denoted as $\Delta_I$.


\begin{theorem}\label{InsertCorrectness}
Algorithm \ref{Alg:SuperiorEdgeInsert} can update the core numbers of vertices after inserting an edge set $E'$ in $O(\Delta_I* \max_{S\subseteq E'}\{m_S + L_S*n_S\})$ time.
\end{theorem}
\begin{IEEEproof}
The algorithm is executed in iterations, and each iteration includes two parts. The first part computes the superior edge set from unprocessed edges in $E'$ by executing Algorithm \ref{Alg:FindKSuperiorEdges} in parallel, and then inserts the computed superior edge set into graph $G$. By Lemma \ref{Them:superioredgesetinsert}, after inserting a superior edge set into the graph, each vertex can increase its core by at most 1. 

After that, in the second part, we identifiy vertices that will increase core numbers by executing Algorithm \ref{Alg:insertK} in parallel. Different processes deal with distinct $k$-superior edge sets, and visit vertices with distinct core numbers. For each inserted $k$-superior edge set $E_k$, we conduct two kinds of operations: (1) a positive DFS that visits vertices in the exKPT, and (2) a negative DFS that will remove vertices that are confirmed not to increase core numbers. By Lemma \ref{le:exkpaths}, visiting vertices in the exKPT is enough to find all vertices whose core numbers potentially increase caused by the insertion of $E_k$. And by Lemma~\ref{corollary:csd}, if a vertex $w$ satisfies $cd[w] \le k$, $w$ will not increase its core number, and it will affect the potential of its neighbors to increase their core numbers. Notice that this influence procedure should be spread across vertices in \textbf{$KPT_w$}, which is done by the negative DFS procedure. When all eges in $E_k$ are handled, the potential vertices are visited and the ones that cannot increase core numbers are removed. All above ensures the corretness of the algorithm.  

As for the time complexity, because in each iteration, for each vertex, at least one inserted edge connected to it can be selected into the superior edge set and processed, there are at most $O(\Delta_I)$ iterations in the algorithm execution. We next consider the time used in each iteration.

Now consider an iteration $i$, and we denote the graph obtained after iteration $i-1$ is $G_i$. Denote by $S$ the superior edge set computed in iteration $i$. The computation of $CSD$ values for vertices in exKPT of $S$ takes $O(m_{S})$ time. The positive DFS visits each vertex in exKPT for one time. Hence the positive DFS procedure takes $n_S$ time. For the negative DFS procedures, notice that after each DFS procedure, if a vertex $v$ is visited, $cd(v)$ is decreased by 1. Hence, each vertex can be visited by at most $L_S$ times, since a vertex will be removed if its $cd$ values is decreased to its core number. Combining together, the total time for an iteration is $O(m_S+L_S*n_S)$.

By above, it can be got the time complexity of the algorithm as stated in the Theorem.  
\end{IEEEproof}

\section{Decremental Core Maintenance}\label{sec:de}
The decremental algorithm is showed in Algorithm \ref{Alg:SuperiorEdgeDelete}. Similar with the incremental algorithm, we deal with deleted edges in iterations. In each iteration, a superior edge set is found using a parallel approach. After that, the graph is updated by deleting the computed superior edge set and the $k$-superior edge sets are assigned to child processes. In each child process, the edges in a $k$-superior edge set is handled one by one similarly. The main difference is that we use $SD$ values to evaluate if a vertex will decrease its core number, and only execute the negative DFS to remove vertices that cannot be in the current $k$-core. When deleting an edge $<u,v>$ with core($u$) $\le$ core($v$), it is checked if $u$ still has enough Superior Neighbors that can help it keep the core number. If $core(u)$ is decreased, Algorithm~\ref{Alg:DeleteRemove} is executed to remove it and disseminate the influence. 

\begin{algorithm}  [htb]\label{Alg:SuperiorEdgeDelete} 
\caption{SuperiorEdgeDelete($G, E', V', core()$)} 
\textbf{Input}\\
The graph, $G=(V,E)$;\\
The deleted edge set, $E'$;\\
The set of vertices $V'$ connected to edges in $E'$\;
The core number $core(v)$ of each vertex in $V$\;
\While{$E'$ is not empty}{
    Let $\mathcal{C}$ be an empty core set\;
\nl \For{each vertex $u$ in $V'$}{
\nl   \If{$u$ connects a superior edge in $E'$ and $core(u)\notin \mathcal{C}$}{
     add $core(u)$ to $\mathcal{C}$\;
        }
    }
\nl \For{each core number $k$ in $\mathcal{C}$ in parallel} {
    $E_k\gets${ComputeSuperiorEdgeSet($k$)}\;
}
\nl delete $\cup_{k\in\mathcal{C}}E_k$ from $G$\; 
\nl delete $\cup_{k\in\mathcal{C}}E_k$ from $E'$\;
\nl     \For{each core number $k$ in $\mathcal{C}$ in parallel}{
     $V_k\gets${$K$-SuperiorDelete($G$,$E_k$,$core()$)}\;
    }
\nl \For{each vertex $v$ in $\cup_{k\in\mathcal{C}}V_k$} {
        $core(v) \gets{core(v)-1}$\;     
    }
}
\end{algorithm} 

\begin{algorithm}  [htb]\label{Alg:deleteK}
\caption{$K$-SuperiorDelete($G, E_k,$ core())}      
\textbf{Input}\\
The graph, $G=(V,E)$;\\
The $k$-superior edge set, $E_k$;\\
The core number $core(v)$ of each vertex in $V$\;
\textbf{Initially}, {$S\gets$ empty stack}\;
for each vertex $v\in V$, $visited[v]\gets{false}, removed[v]\gets{false}, cd[v]\gets{0}$\;

\nl \For{each $e_i=<u,v> \in E_k$}{  
\nl \lIf{$core(u)\geq core(v)$}{
$r\gets{v}$}
\lElse{$r\gets{u}$}
\nl \If{core($v$) $\neq$ core($u$)}{
        \If{visited[$r$] = false}{
            $visited[r] \gets true$\;
            $cd[r] \gets SD(r)$\;
        }
\nl         \If{removed[$r$] = false}{
            \If{cd[$r$] $< k$}{
\nl                DeleteRemove($G,core()$,\\$cd[],removed[],k,r$)
            }
        }        
    }
\nl    \Else{
\nl        \If{visited[u] = false}{
                $visited[u] \gets {true}$\;
                $cd[u] \gets{ SD[u]}$\;
            }
\nl        \If{removed[$u$] = false}{
\nl            \If{cd[$u$] $< k$}{
                 DeleteRemove($G,core()$,$cd[],removed[],k,u$)
                }
            } 
\nl         \If{visited[$v$] = false}{
                $visited[v] \gets{true}$\;
                $cd[v] \gets{SD(v)}$\;
            }
\nl         \If{removed[$v$] = false}{
                \If{cd[$v$] $< k$}{                 
\nl                DeleteRemove($G,core()$,$cd[],removed[],k,v$)
                }
            } 
        }
    }
\nl \For{each vertex $v$ in $G$}{
    \If{removed[$v$] = true and visited[$v$] = true}{
         $V_c \gets{ V_c\cup\{v\}}$ }
    }

\nl \textbf{return} $V_c$;
\end{algorithm}  

\begin{algorithm}[htb]\label{Alg:DeleteRemove}
\caption{DeleteRemove($G$,core(),cd[],removed[],$k,r$)}  
\nl {$S\gets{empty\ stack}$}\;
\nl $S.push(r)$\;
\nl $removed[r] \gets{true}$\;
\nl \While{S is not empty}{ 
        $v\gets{S.pop()}$\;
        \For{each $<v,w>\in E$}{
\nl         \If{core($w$) = $k$}{
\nl             \If{visited($w$) = false}{
                    $visited[w] \gets{true}$\;
                    $cd[w] \gets{cd[w] + SD(w)}$\;
                }
\nl             $cd[w]\gets{cd[w]-1}$\;
                \If{cd[$w$] $<$ $k$ and removed[$w$] = false}{
\nl                $S.push(w)$\;
\nl                 $removed[w] \gets {true}$\;
                }        
            }
        } 
    }
\end{algorithm}

\textbf{Performance Analysis.}
We next analyze the correctness and efficiency of the proposed decremental algorithm. At first, some notations are defined, which will be used in measuring the time complexity of the algorithm.

For graph $G=(V,E)$, the deleted edge set $E'$ and a subset $R$ of $E'$, let $G_R=(V,E\setminus R)$ and $K(G_R)$ be the set of core numbers of vertices in $G_R$. 

For $G_R$, let $F_R = \max_{u\in V}\{SD(u)-core_{G_R}(u),0\}$.

For $k\in K(G_R)$, let $V_R(k)$ be the set of vertices with core number $k$ and $n_R=\max\{|V_R(k)|: k\in K(G_R)\}$.

Denote by $E(V_R(k))$ the set of edges connected to vertices in $V_R(k)$. We then define $m_R$ as follows, 
\begin{equation*}
m_R=\max_{k\in K(G_R)}\{|E(V_R(k))|\}.
\end{equation*}
$F_R$, $n_R$ and $m_R$ will depict the time used in each iteration in the algorithm execution. Furthermore, we define the \emph{maximum deleted degree} as the maximum number of edges deleted from each vertex in $V$, denoted as $\Delta_D$. 

Using a similar argument as that for analyzing the incremental algorithm, we can get the following result, which states the correctness and efficiency of the decremental algorithm. The detailed proof is put in Appendix.

\begin{theorem}\label{DeleteCorrectness}
Algorithm \ref{Alg:SuperiorEdgeDelete} can update the core numbers of vertices after inserting an edge set $E'$ in $O(\Delta_D*\max_{R\in E'}\{(m_R+F_R*n_R)\})$ time.
\end{theorem}

\section{Experiment Studies}\label{sec:experiment}
In this section, we conduct empirical studies to evaluate the performances of our proposed algorithms. The experiments use three synthetic datasets and seven real-world graphs, as shown in Table \ref{table_graph}. 

There are two main variations in our experiments, the original graph and the inserted/deleted edge set. We first evaluate the efficiency of our algorithms on real-world graphs, by changing the size and core number distribution of inserted/deleted edges. Then we evaluate the scalability of our algorithms using synthetic graphs, by keeping the inserted/deleted edge set stable and changing the sizes of synthetic graphs. At last, we compare our algorithms with the state-of-the-art core maintenance algorithms for single edge insertion/deletion, TRAVERSAL algorithms given in \cite{Sar2016Incremental}, to evaluate the acceleration ratio of our parallel algorithms. The comparison experiments are conducted on four typical real-world datasets.


All experiments are conducted on a Linux machine with Intel Xeon CPU E5-2670@2.60GHz and 64 GB main memory, implemented in C++ and compiled by g++ compiler. 

\textbf{Datasets.} We use seven real-world graphs and random graphs generated by three models. The seven real-world graphs can be downloaded from SNAP \cite{SNAP}, including social network graphs (LiveJournal, Youtube, soc-Slashdot), collaboration network graphs (DBLP, ca-astroph), communication network graphs (WikiTalk) and Web graphs (web-BerkStan). The synthetic graphs are generated by the SNAP system using the following three models: the Erd\"os-R\textbf{$\acute{e}$}nyi (ER) graph model \cite{ER}, which generates a random graph; the Barabasi-Albert (BA) preferential attachment model \cite{BA}, in which each node creates $k$ preferentially attached edges; and the R-MAT (RM) graph model \cite{RM}, which can generate large-scale realistic graphs similar to social networks. For all generated graphs, the average degree is fixed to 8, such that when the number of vertices in the generated graphs is the same, the number of edges is the same as well.

\begin{figure}
 \subfigure[Real-world Graphs]{
    \label{core1} 
    \includegraphics[width=1.6in]{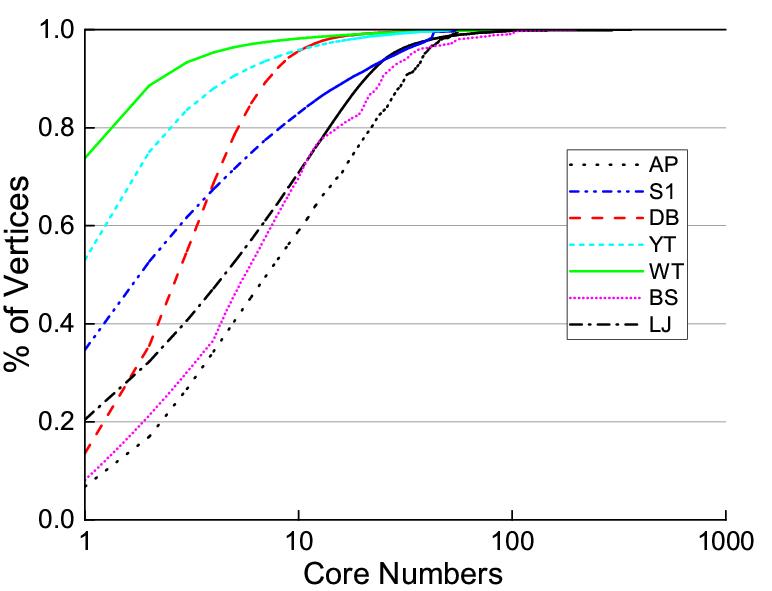}}
 \subfigure[Generated Graphs]{
    \label{core2} 
    \includegraphics[width=1.6in]{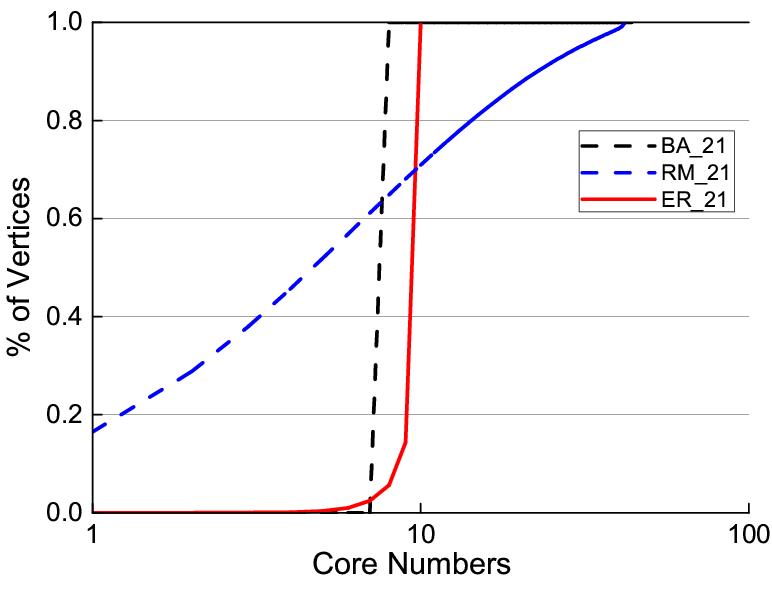}}
 \caption{Core Distribution}
 \label{cores} 
\end{figure}

Fig. \ref{core1} and Fig. \ref{core2} show the core number distributions of the seven real-world graphs and the generated graphs with $2^{21}$ vertices. From Fig. \ref{core1}, it can be seen that in real-world graphs, more than 60 percent of vertices have core numbers smaller than 10. Especially, in WT (wiki-Talk), more than 70\% of vertices have core number 1. For the core distributions of generated graphs, as shown in Fig. \ref{core2}, in the BA graphs, all vertices have a core number of 8. In the ER graph, the core numbers of vertices are small and the max core number of vertices is 10, but almost all vertices have core numbers close to the max one. The RM graph are more close to real-world graphs, where most vertices have small core numbers and as the core number $k$ increases, the percentage of vertices with core number $k$ decreases. As shown later, the core distribution of a graph will affect the performances of our algorithms. 


The \emph{core number} of an edge is defined as the smaller core number of its two endpoints. We use the \emph{average processing time per edge} as the efficiency measurement of the algorithms, such that the efficiency of the algorithms can be compared in different cases.

\begin{table}[!t]
\renewcommand{\arraystretch}{1.3}
\caption{Real-world graph datasets}
\label{table_graph}
\centering
\begin{tabular}{|c|c|c|c|c|}
\hline
Datasets & n=$|V|$ & m=$|E|$ & max degree & max core\\
\hline
AP(ca-Astroph)  & 18.7K     & 198.1K & 504      & 56  \\  
S1(soc-Slashdot)& 82.1K     & 500.5K & 2548     & 54  \\    
DB(DBLP)        & 0.31M     & 1.01M  & 343      & 113 \\  
YT(YouTube)     & 1.13M     & 1.59M  & 28754    & 35    \\
WT(wiki-Talk)   & 2.4M      & 9.3M   & 100029   & 131   \\
BS(web-BerkStan)& 0.68M     & 13.3M  & 84230    & 201   \\
LJ(LiveJournal) & 4.0M      & 34.7M  & 20334    & 360   \\
\hline
\end{tabular}
\end{table}


\subsection{Performance Evaluation}

We evaluate the impacts of three factors on the algorithm performance: the size of inserted/deleted edges, the core number distribution of edges inserted/deleted, and the original graph size. The first factor affects the iterations needed to process the inserted/deleted edges, and the last two factors affect the processing time in each iteration. The first two evaluations are conducted on real-world graphs, and the third one is on synthetic graphs. 

\begin{figure}
 \subfigure[Insertion]{
    \label{Size_ins} 
    \includegraphics[width=1.6in]{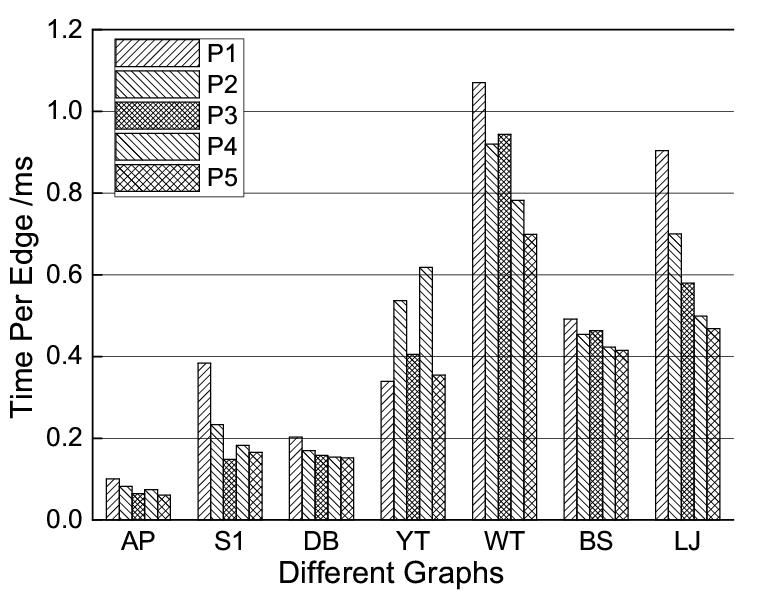}}
 \subfigure[Deletion]{
    \label{Size_del} 
    \includegraphics[width=1.6in]{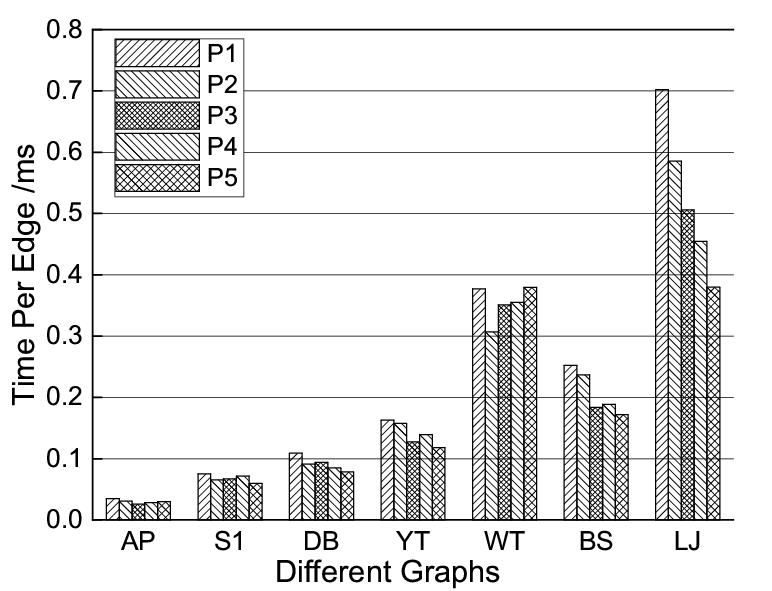}}
 \caption{Impact of Inserted/Deleted Edge Number}
 \label{SizeChange} 
\end{figure}

We first evaluate the impact of the number of inserted/ deleted edges on the performances of our algorithms. The results for the incremental and decremental maintenance algorithm are illustrated in Fig. \ref{Size_ins} and Fig. \ref{Size_del} respectively. In the experiments, we randomly insert/delete $P_i$\% edges with respect to the original graph, where $P_i=3*i$ for $i=1,2,3,4,5$. In Fig. \ref{Size_ins} and Fig. \ref{Size_del}, the x-axis represents the datasets, and the y-axis represents the average processing time per edge. It can be seen that the processing time per edge is less than $1.2ms$ in all cases, and except for WT and LJ, the processing time is much smaller than $1.2ms$. The figures show that the processing time decreases as the number of inserted/deleted edges increases, which demonstrates that our algorithms are suitble for handling large amount of edge insertions/deletions. In this case, more edges can be selected into the superior edge set in each iteration, and hence our algorithms achieve better parallelism. Furthermore, Fig. \ref{Size_ins} and Fig. \ref{Size_del} also illustrate that it needs a larger average processing time when the size of original graphs increases. The only exception is the WT graph. Though the graph has a smaller size than BS and LJ graphs, the average processing time is larger. This is because the core distribution of WT is rather unbalancing, as showed in Figure \ref{core1}, where most vertices possess the same core number. In this extremal case, on the one hand, each iteration in the algorithm takes more time in processing the inserted edges, as more vertices need to be traversed, and on the other hand, the parallelism of the algorithm is very limited, as most edges are inserted to vertices with the same core.

\begin{figure}
 \subfigure[Insertion]{
    \label{core_ins} 
    \includegraphics[width=1.6in]{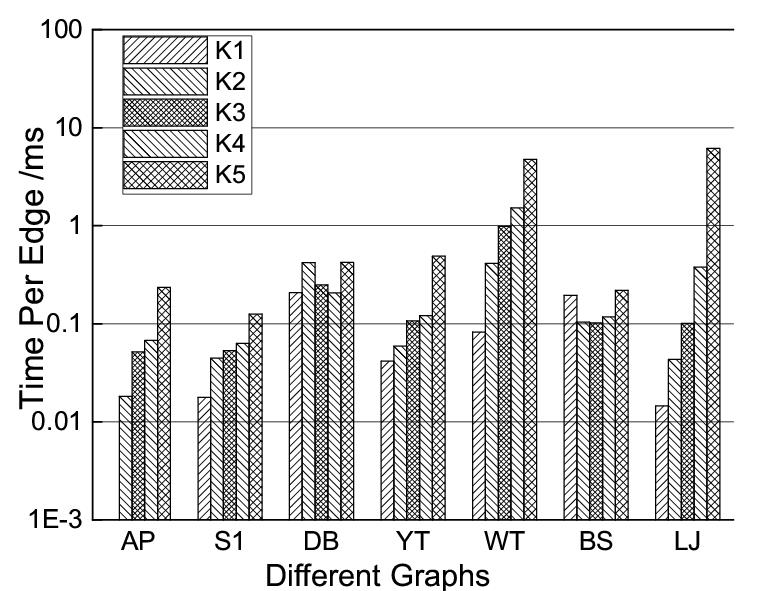}}
 \subfigure[Deletion]{
    \label{core_del} 
    \includegraphics[width=1.6in]{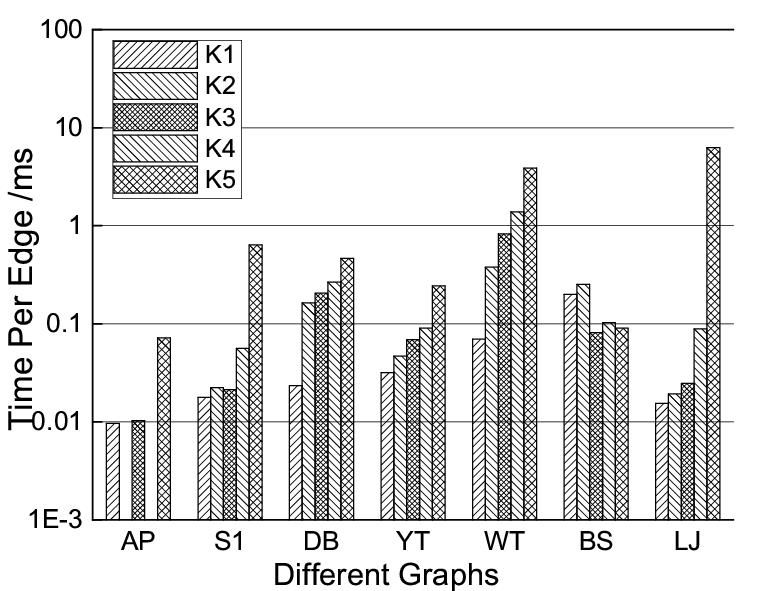}}
 \caption{Impact of Core Number of Inserted/Deleted Edges}
 \label{coreChange} 
\end{figure}

We then evaluate the impact of the core number distribution of inserted/deleted edges on the algorithm performance. The results are illustrated in Fig. \ref{coreChange}. In particular, by the core distributions showed in Fig. \ref{core1}, we choose five typical core numbers \{$K1,K2,K3,K4,K5$\} in an increasing order for each of the seven graphs. For each core number, 20\% edges of that core number are selected randomly as the update edge set. From Fig. \ref{coreChange}, it can be seen that larger core number induces a larger average processing time. This is because, when inserting/deleting edges to vertices with larger core numbers, the degree of these vertices generated by these inserted edges is larger. In our algorithm, only one superior edge can be handled for each vertex in each iteration. Hence, it takes more iterations to process the inserted/deleted edges. But on the other hand, it can be also seen that the processing time per edge does not vary significantly.

\begin{figure}
 \subfigure[Insertion]{
    \label{graph_ins} 
    \includegraphics[width=1.6in]{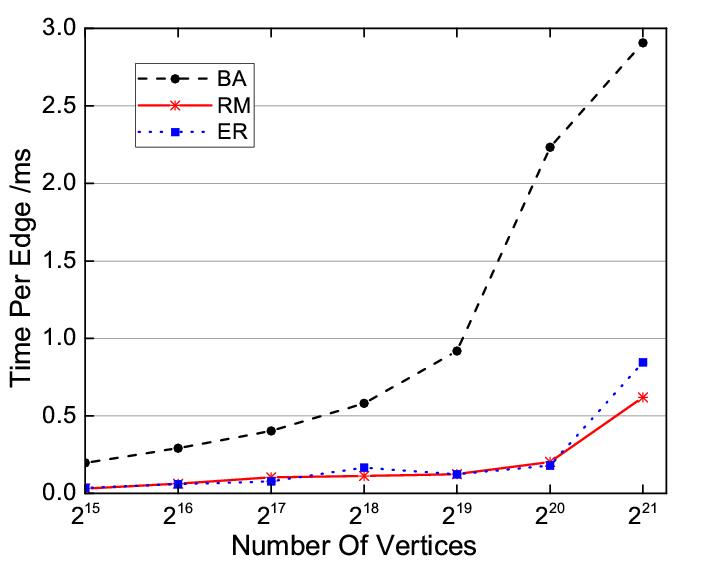}}
 \subfigure[Deletion]{
    \label{graph_del} 
    \includegraphics[width=1.6in]{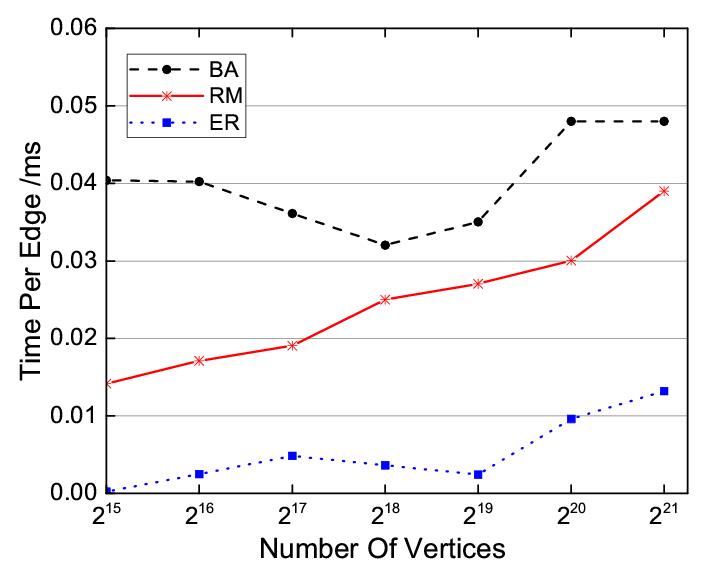}}
 \caption{Impact of Orginal Graph Size}
 \label{graphChange} 
\end{figure}

We finally evaluate scalability of our algorithms in synthetic graphs, by letting the number of vertices scale from $2^{15}$ to $2^{21}$ and keeping the average degree fixed as 8. The results are shown in Fig. \ref{graphChange}. In the experiments, for each graph, we randomly select 10000 edges as the update set. In Fig. \ref{graphChange}, the x-axis represents the number of vertices in the graph, and the y-axis represents the average processing time per edge. Fig. \ref{graphChange} shows that though the graph size increases exponentially, the average processing time increases linearly. It demonstrates that our algorithms can work well in graphs with extremely large size. From the figures, it can be also seen that the processing time in the BA graph is larger than those of the other two graphs. This is because all vertices in the BA graph have the same core number 8. This means that in our algorithm, all edges are initially handled in one process, and hence the parallelism is poor in this extreme case. This can be seen as the worst case for our algorithms. However, as shown in Fig. \ref{core1} and Fig. \ref{core2}, real-word graphs exhibit much better balance in core number distribution.


\begin{figure}
 \subfigure[Insertion]{
    \label{cmp_ins} 
    \includegraphics[width=1.6in]{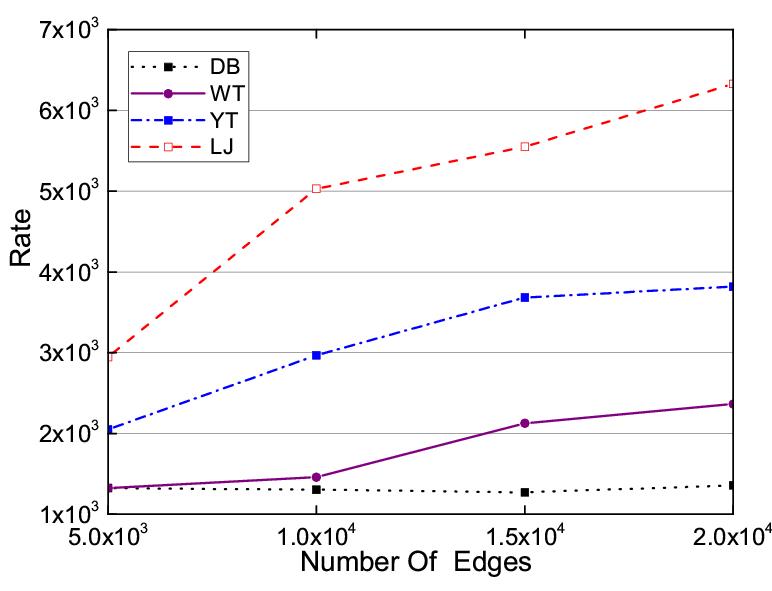}}
 \subfigure[Deletion]{
    \label{cmp_del} 
    \includegraphics[width=1.6in]{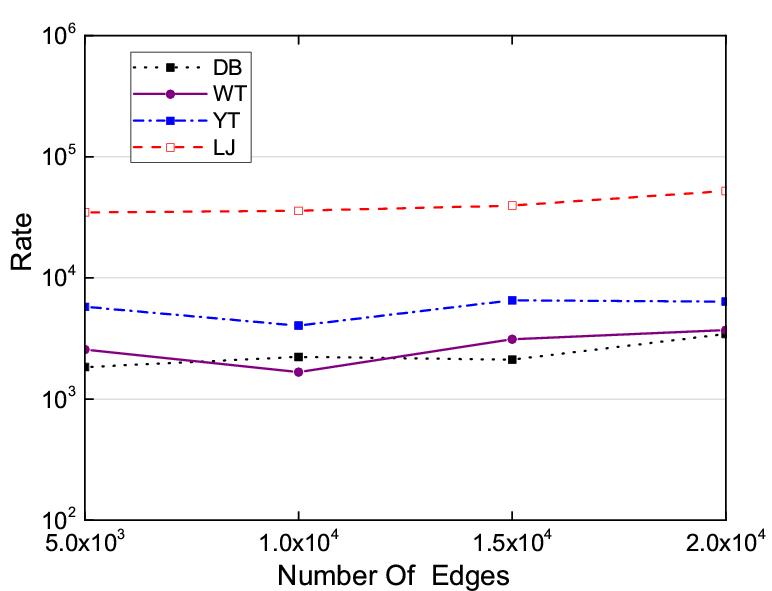}}
 \caption{Comparison with the TRAVERSAL Algorithm}
 \label{Comparison} 
\end{figure}
\subsection{Performance comparison}
In this section, we evaluate the acceleration ratio of our parallel algorithms, comparing with algorithms sequentially handling edge insertions/deletions. We compare with the state-of-the-art sequential algorithm, \textbf{TRAVERSAL} algorithms given in \cite{Sar2016Incremental}. The comparison is conducted on four typical real-world graphs, DB, WT, YT and LJ in Table \ref{table_graph}. For each graph, we randomly select 5K-20K edges as the update set. The evaluation results are illustrated in Fig. \ref{cmp_ins} and Fig. \ref{cmp_del} respectively. In the figures, the x-axis and y-axis represent the number of inserted/deleted edges and the acceleration ratio, respectively.

From Fig. \ref{cmp_ins} and Fig. \ref{cmp_del}, it shows that in almost all cases, our algorithms achieves an acceleration ratio as large as $10^3$ times in both incremental and decremental core maintenance. The acceleration ratio increases as the number of edges inserted/deleted increases, which illustrates that our algorithms have better parallelism in scenarios of large amounts of graph changes. Furthermore, it is also shown that our algorithms achieve larger acceleration ratios as the graph size increases.



All evaluation results show that our algorithms exhibit good parallelism in core maintenance of dynamic graphs, comparing with sequential algorithms. The experiments illustrate that our algorithms are suitable for handling large amounts of edge insertions/deletions in large-scale graphs, which is desirable in realistic implementations.

\section{Conclusion}\label{sec:conclusion}
In this paper, we present the first known parallel algorithms for core maintenance in dynamic algorithms. Our algorithms have significant accelerations comparing with sequential processing algorithms that handle inserted/deleted edges sequentially, and reduce the number of iterations for handling $s$ inserted/deleted edges from $s$ to the maximum number of edges inserted to/deleted from a vertex. Experiments on real-world and synthetic graphs illustrate that our algorithms implement well in reality, especially in scenarios of large-scale graphs and large amounts of edge insertions/deletions.

For the future work, it deserves more efforts to discovering structures other than superior edge set that can help design parallel core maintenance algorithms. Furthermore, it is also meaningful to design parallel algorithms for maintaining other fundamental vertex parameters, such as betweenness centrality~\cite{HAY15}.

\begin{appendix}
\textbf{Proof of Theorem \ref{DeleteCorrectness}.}
\\

The deletion algorithm is executed in iterations, and each iteration includes two parts. The first part is similarly as the insertion case, which computes the superior edge set in parallel, and then deletes the computed superior edge set from graph $G$. By Lemma \ref{Them:superioredgesetdelete}, after deleting such a superior edge set $E_k$ from the graph, each vertex can decrease its core by at most 1.

Then in the second part, we identifiy vertices that will decrease core numbers by executing Algorithm \ref{Alg:deleteK} in parallel. In each child process, it is sufficient to visit vertices in the exKPT of $E_k$ to find all vertices whose core numbers may decrease according to Lemma \ref{le:exkpaths}. For each edge, we start a negative DFS to remove vertices that are confirmed to decrease core numbers. And by Lemma~\ref{corollary:sd}, for a vertex $v$, if $SD[v] \le k$, $v$ will decrease its core number, and this will affect the $SD$ value of its neighbors. So we use a variable value $cd$ to represent the dynamic changes of $SD$ value. After all edges are handled, vertices in exKPT are visited and the ones that cannot be in the current $k$-core are marked as removed. All above ensures the corretness of the algorithm.  

As for the time complexity, the iterations needed is similarly bounded by $O(\Delta_D)$ as the insertion case. We next consider the time used in each iteration.

Now consider an iteration $i$, denote the superior edge set computed in current iteration $i$ as $R$. The computation of $SD$ values for vertices in exKPT of $R$ takes $O(m_R)$ time. For the negative DFS procedures, if a vertex $v$ is visited, $cd(v)$ is decreased by 1. Hence, each vertex can be visited by at most $F_R$ times, since a vertex will be removed if its $cd$ values is decreased below to its core number. Combining together, the total time for an iteration is $O(m_R+F_R*n_R)$.

By above, it can be got the time complexity of the algorithm as stated in the Theorem.
\end{appendix}
\end{document}